\documentclass[useAMS,usenatbib,usegraphicx]{mn2e}
\usepackage{graphicx}                                            
\usepackage{times}
\usepackage{float}
\usepackage{rotating}
\usepackage{epstopdf}
\usepackage{multirow}

\def\ltsima{$\; \buildrel < \over \sim \;$}
\def\simlt{\lower.5ex\hbox{\ltsima}}
\def\gtsima{$\; \buildrel > \over \sim \;$}
\def\simgt{\lower.5ex\hbox{\gtsima}}
\def\gsimeq
{\hbox{\raise0.5ex\hbox{$>\lower1.06ex\hbox{$\kern-1.07em{\sim}$}$}}}
\def\lsimeq
{\hbox{\raise0.5ex\hbox{$<\lower1.06ex\hbox{$\kern-1.07em{\sim}$}$}}}

\def\xmm{{\it XMM-Newton }}

\def\xmm{{\it XMM-Newton}}
\def\chandra{{\it Chandra}}
\def\suzaku{{\it Suzaku}}
\def\rxte{{\it RXTE}}

\def\apj{ApJ}
\def\mnras{MNRAS}
\def\aap{A\&A}
\def\apjl{ApJ}

\def\nat{Nature}
\def\iaucirc{IAU~circular}

\def\exo{EXO~0748-676}

\def\Fevc{Fe {\sc xxv}}
\def\Fevs{Fe {\sc xxvi}}

\def\xis{XIS}
\def\xis1{XIS1}
\def\xis2{XIS2}
\def\xis3{XIS3}

\title[] 
 {{A connection between accretion state and {\it Fe~K absorption} in an accreting
neutron star: black hole-like soft state winds?}}

 \author[G.\ Ponti et al. ]
 {Gabriele~Ponti$^{1}$, Teodoro~Mu\~noz-Darias$^{2}$ and Robert~P.~Fender$^{2}$
\\
   $^1$ Max Planck Institute f{\"u}r Extraterrestriche Physik,  Giessenbachstrasse, D-85748, Garching, Germany\\
   $^2$ University of Oxford, Department of Physics, Astrophysics, Denys Wilkinson 
   Building, Keble Road, Oxford OX1 3RH, UK\\
}
\pagerange{\pageref{firstpage}--\pageref{lastpage}}
\usepackage{times}
\begin{document}
\label{firstpage}
 \maketitle
\begin{abstract}
High resolution X-ray spectra of accreting stellar mass Black Holes reveal the presence 
of accretion disc winds, traced by high ionisation Fe~K lines. 
These winds appear to have an equatorial geometry and to be observed only during 
disc dominated states in which the radio jet is absent. 
Accreting neutron star systems also show equatorial high ionisation 
absorbers. However the presence of any correlation with the accretion 
state has not been previously tested. We have studied EXO~0748-676, a transient 
neutron star system, for which we can reliably determine the 
accretion state, in order to investigate the Fe~K absorption/accretion state/jet connection. 
Not one of twenty X-ray spectra obtained in the hard state revealed
any significant Fe~K absorption line. However,
intense \Fevc\ and \Fevs\ 
(as well as a rarely observed Fe~{\sc xxiii} line plus S~{\sc xvi}; a blend of S~{\sc xvi} 
and Ar~{\sc xvii}; Ca~{\sc xx} and Ca~{\sc xix}, possibly produced by the same high 
ionisation material) absorption lines ($EW_{\rm Fe~{\sc xxiii-xxv}}=31\pm3$, 
$EW_{\rm \Fevs}=8\pm3$~eV) are clearly detected during the only soft state observation. 
This suggests that the connection between Fe~K absorption and states (and 
anticorrelation between the presence of Fe~K absorption and jets) is also valid 
for \exo\ and therefore it is not a unique property of black hole systems but a more 
general characteristic of accreting sources. 
\end{abstract}


\section{Introduction}

The advent of the new generation of X-ray telescopes yielded a burst of detections 
of highly ionised absorption Fe features (e.g. \Fevc\ and \Fevs) in low mass X-ray 
binaries harbouring both black holes (BH) and neutron stars (NS)  (Brandt et al. 2000; 
Lee et al. 2002; Parmar et al. 2002; Jimenez-Garate et al. 2003; Boirin et al. 2003; 
2004; 2005; Ueda et al. 2004; Miller et al. 2006a, b). 
In NS, at first, it has been realised that such features seems to be detected 
only in dipping - high inclination - sources (Diaz-Trigo et al. 2006). 
The same has also proven to be true in BH systems (Ponti et al. 2012), 
therefore implying an equatorial geometry of these absorbers. 

More recently, high resolution observations of BH systems showed that 
these absorption features have significant outflow velocities, and therefore 
are thought to be the signature of equatorial winds. The estimated wind mass 
outflow rates are generally of the order or higher than the 
mass accretion rates (Lee et al. 2002; Ueda et al. 2004; Neilsen et al. 2011; 
Ponti et al. 2012), suggesting that these winds are a fundamental ingredient 
in the accretion process. 
Another key aspect is that they are observed primarily in the 
so-called soft states (see Belloni et al. 2011 for a recent review on X-ray states), 
when the accretion flow can be, at least, partially described 
by an optically thick, geometrically thin disc (Shakura \& Sunyaev 1973) and 
the radio jet is quenched (Fender et al. 2004). 
Consequently, these winds are not observed during hard states, 
characterised by a strong Comptonisation component and stable radio emission 
from a compact jet. These observational facts suggest a deep link between 
the presence of an equatorial disc wind and the accretion disc state and/or 
the jet (Neilsen \& Lee 2009; Ponti et al. 2012). 

NS systems are also known to display several X-ray states. In particular, 
when accreting at low to moderate rates ($0.01-0.5$~L$_{\rm Edd}$), 
these are in many aspects analogous to the hard and soft 
states observed in BH (see van der Klis 2006). 
They are also observed to alternate following marked hysteresis 
patterns (Mu\~noz-Darias et al. 2014) and to be similarly connected with 
the presence/absence of the (radio) jet (Migliari \& Fender 2006).  
Although the Fe~K absorbers in NS show similar properties to
BH systems (e.g. Fe~K lines equivalent widths, 
absorber ionisation states, equatorial geometries; Diaz-Trigo \& 
Boirin 2013), an outstanding question is whether the same state-wind 
connection applies for accreting neutron stars too. 
To test this, high quality observations of a high inclination NS system showing 
both hard and soft states are needed. 

\exo\ (=UY Vol) is a neutron star low mass X-ray binary discovered in outburst 
by EXOSAT in 1985 (Parmar et al. 1985). 
It was active for 23 years, until it finally returned to quiescence in 2008 
(Hynes \& Jones 2008). The system showed both absorption dips and 
eclipses implying a high orbital inclination. Assuming a primary mass of 
$M_{\rm NS}\sim1.4$~M$_{\odot}$ (see Mu\~noz-Darias et al. 2009 and 
Ratti et al. 2012 for dynamical studies) an inclination of $75<i<83^{\circ}$ 
was estimated by modelling the X-ray light-curve (Parmar et al. 1985, 1986).

\begin{figure*}
\includegraphics[width=0.9\textwidth,height=0.55\textwidth]{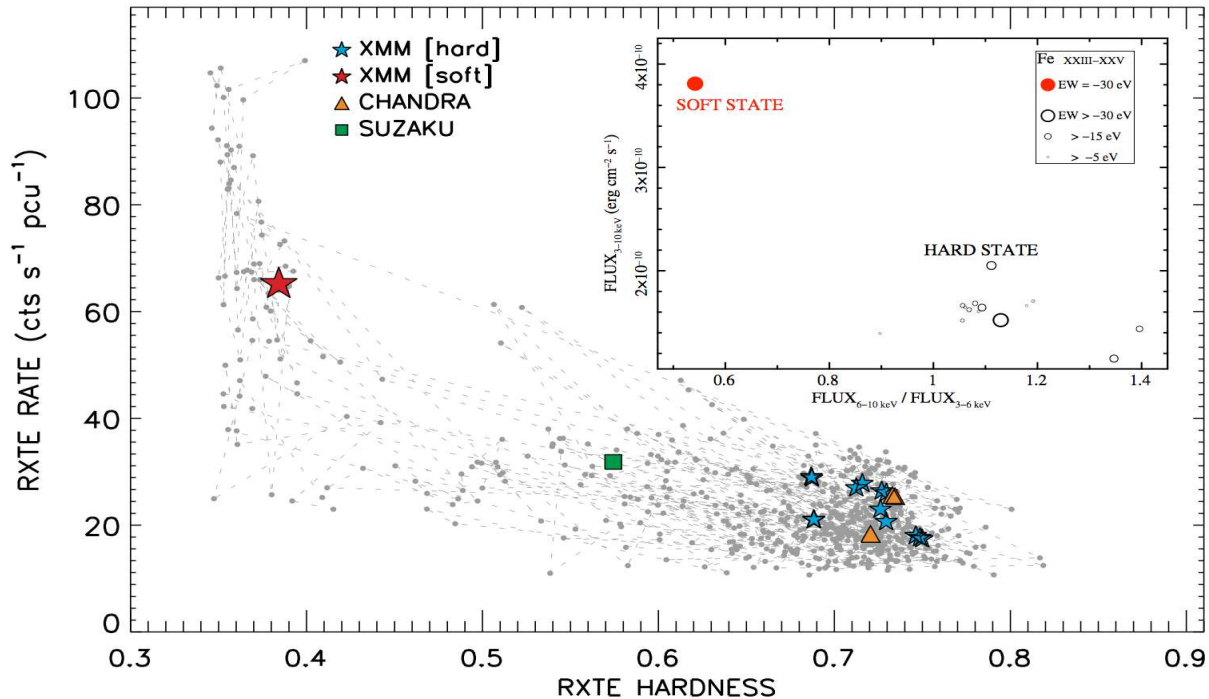}
\caption{Hardness intensity diagram of \exo\ obtained with \rxte\ (grey dots). The dashed lines join contiguous observations. The (RXTE) hardness and count-rate values at the times of the \suzaku, \chandra\ and \xmm\ hard state observations are indicated as a green square, orange triangles and blue stars, respectively. The red star indicates to the only soft state \xmm\ observation. \textbf{Inset:} hardness intensity diagram obtained directly from the \xmm, \chandra\ and \suzaku\ observations presented in 
this work (i.e non \rxte). Detections of Fe~K absorption lines (Fe~{\sc xxiii-xxv} in this case) 
are shown with red filled symbols, while upper limits are shown with empty circles. 
The symbol size indicate the equivalent width of the line 
detected/upper limit. During none of the hard state observations Fe~K absorption 
is observed, while strong Fe~{\sc xxiii-xxv} and \Fevs\ are detected in the soft state one.}
\label{HID}
\end{figure*}

\exo\ shows a variety of absorption and emission components. 
The strongest are neutral and photo-ionised absorption features 
(Jimenez-Garate et al. 2003; Diaz-Trigo et al. 2006; van Peet et al. 2009), 
with their column densities increasing and ionisation parameters 
decreasing during dips. In addition, collisionally ionised absorbers and Oxygen 
emission lines have also been detected (van Peet et al. 2009).

\section{Observations and data reduction}

%\subsection{\xmm}
\label{xmmDR}

\textbf{\textit{XMM-Newton:}} 
We use archival \xmm\ data starting from the Observation Data Files (ODF). 
They were processed with the MPE development version 13.5.0 of the \xmm\ 
Science Analysis System (SAS), applying the most recent calibrations. 
For each spectrum, the response matrix and effective 
area has been computed with the tasks {\sc rmfgen} and {\sc arfgen}. 
Because of the higher effective area in the Fe~K band, we use only the data from 
the EPIC-pn camera. Up to 14-02-2013 there are 21 observations, publicly 
available in the \xmm\ archive (Tab. \ref{data}), pointed at \exo\ and with EPIC-pn 
clean exposure longer than 2~ks. We summed the spectra and response matrices of 
the consecutive short (3-5~ks) observations accumulated during 
revolution 212 to increase the signal-to-noise. 

Several observations in imaging mode were 
affected by photon pile-up (Tab. \ref{data}). Whenever significant pile-up 
is detected, we use an annular extraction region centred on 
the source, with inner radius of $r_{\rm in}=9.25''$ and outer radius of 
$r_{\rm out}=45''$ (see e.g. van Peet et al. 2009), otherwise a circular 
region with $45''$ radius is used. The background was selected 
from a region of similar size and shape and on the same detector chip 
as the source region. 

In order to identify and remove type I bursts from the analysis we used a 3~s resolution 
hard X-ray light curve ($5\leq E\leq10$ keV)
since this energy band is only marginally affected by dipping (Diaz-Trigo et 
al. 2006; van Peet et al. 2009). 
In the same way, but with 15~s time bins, we selected the eclipses starting 
and ending times. The thresholds applied are reported in Tab. \ref{data}. 
We then identified the periods of enhanced particle activity by calculating the 
full detector light curve (once excluded the events within a 2.5~arcmin 
region from the target) in the 12-15~keV band. 
Intervals with count rate higher than the threshold 
(varying according to the observing mode and source 
brightness) specified in Tab. \ref{data} were consequently filtered out. 
Finally, to separate dipping and persistent emission, we divide the 
$5-10$~keV light curve by the $0.5-5$~keV one, since absorption dips 
are revealed by sudden increases in the hardness ratio. 
Following van Peet et al. (2009) we determined the average hardness 
ratio of the intervals clearly belonging to the persistent emission and 
selected as \textit{dipping} the periods with hardness ratios 1.5 times larger 
than the persistent value. 
After applying the particle background cut and the removal of bursts
and eclipses, we extracted, for each observation, a source and 
background spectrum for both the persistent and dipping periods 
(see Tab. \ref{data}). 

For the only observation in timing mode (Tab. \ref{data}), the source 
photons were selected from a region within {\sc rawx} = 22 and 
{\sc rawx} = 54 and background photons within {\sc rawx} = 1 and 
{\sc rawx} = 17\footnote{For comparison and to investigate further 
the reliability of the energy scale of the EPIC-pn instrument, we also analysed 
the {\sc mos2} data. We selected the source photons within {\sc rawx} = 282 and 331 
and the background photons within {\sc rawx} = 257 and 280.
The same GTI used for the pn are used. We also reduced (and produced 
spectra and response matrices of) the RGS data with the {\sc sas} command {\sc rgsproc}. }. The calibration of the energy scale of the EPIC cameras in timing mode
is difficult. Uncertainties of the order of several $\sim10$~eV can be 
observed (see the \xmm\ Calibration Technical Note 
0083; Guainazzi et al. 2012).
Following the recommendation of the EPIC calibration team, 
we apply the X-ray loading correction (not default 
in version 13.5.0 of the {\sc sas}) to obtain the best possible energy scale. 
We are aware that an uncalibrated energy scale produces spurious features 
at the energies of the mirror edges. However, the effective area above 
2.5~keV, and in particular in the Fe~K band, shows no strong edge.
Therefore, we focus our analysis on the 2.5-10~keV band only.

%\subsection{\chandra}
\textbf{\textit{Chandra:}} 
We reduced the data in a standard manner (see Ponti et al. 2012 for details) 
using version 4.4 of the {\sc ciao} analysis package. We started from the {\it evt1} 
file, accepted only the standard event grades from the nominal good time intervals 
and excluded bad pixels. 
Because of the superior effective area and energy resolution at the Fe~K 
complex energy, we analyse only the {\sc heg} data. 

%\subsection{\suzaku}
\textbf{\textit{Suzaku:}} 
The \suzaku\ {\sc xis} event files were processed using 
the standard pipeline ({\sc aepipeline} version 1.0.1) with the calibration files available 
(2013-01-10 release), using the {\sc ftools} package of {\sc lheasoft} version 
6.13 and adopting the standard filtering criteria.
Source and background spectra were extracted from circular regions ($r=140$"), 
centred on, and away from the source, respectively. Response matrices and 
ancillary response files were produced with {\sc xisrmfgen} and {\sc xissimarfgen} 
tools. Only the data from the {\sc xis0} and {\sc xis3} instruments were used. 
%\subsection{\rxte}

\textbf{\textit{RXTE:}} 
We used all (707) the observations taken by \textit{Rossi X-ray Timing Explorer} 
to monitor the status of the source at a given time (see Fig. \ref{HID}). 
The \rxte-PCA Standard 2 mode (STD2) was used for the production of 
count-rates and colours. It covers the 2--60 keV energy range with 129 channels. 
For each observation net count rate corresponds to STD2 channels 0--31 
(2--15 keV). 
For the state classification (see below) we defined a 
\textit{hardness}\footnote{Ratio of counts between the channels 20--33 
(10--16 keV) and 11--19 (6--10 keV), respectively.} and computed  power 
density spectra (PDS; see Belloni et al. 2006) and the root-mean-square (rms) 
variability following Mu\~noz-Darias et al. (2011).

All the fits were performed using the {\sc Xspec} software (version 12.7.0). 
The errors and upper limits are reported at the 90 per cent confidence level for 
one interesting parameter. 

\section{State classification}
Our \rxte\ analysis of \exo\ is part of a contemporaneous study on 
a large sample of accreting NS for which the long-term X-ray 
behaviour is being investigated in detail using X-ray colours and 
variability (Mu\~noz-Darias et al. 2014). Figure \ref{HID} shows the \rxte\ Hardness Intensity Diagram (HID) using all the data available (one point per observation).
During its 23 years-long outburst \exo\ behaved as a persistent source 
accreting at low-to-moderate rates ($\leq 30$\% L$_{\rm Edd}$). 
It stayed most of the time in the hard state, but displayed a number of 
transitions to brighter and softer states, when the fast variability 
(fractional rms; not shown here) also dropped from $\sim25$ to $\sim5$~\%, as observed in 
other sources of its class (Mu\~noz-Darias et al. 2014) and in black 
hole systems. 
In Fig. \ref{HID} we have indicated the (RXTE) hardness and count-rate values at the times of the \suzaku, \chandra\ and \xmm\ hard state observations as a green square, orange triangles and blue stars, respectively. The red star corresponds to the only soft state \xmm\ observation. 
To do this, we used measurements from simultaneous \rxte\ data (as in the case of the soft observation) or we interpolated the closest in time. In the latter case, we also checked the 1 day \rxte-\textit{All Sky Monitor} 
light-curve to ensure that no flux increase (i.e. suggesting a transition 
to the soft state) occurred at that time. As a definitive proof of the above, 
we show in the inset of Fig. \ref{HID} the HID directly extracted 
from all the \xmm, \chandra\ and \suzaku\ observations included in the 
analysis. The total flux is computed over the 3-10~keV band and the X-ray colour 
displayed as the ratio between the fluxes in the range 6-10~keV and 
the 3-6~keV.

\section{Soft state spectrum}
\label{Sectsoft}
\begin{figure}
\hspace{-1cm}
\includegraphics[width=0.4\textwidth,height=0.58\textwidth,angle=-90]{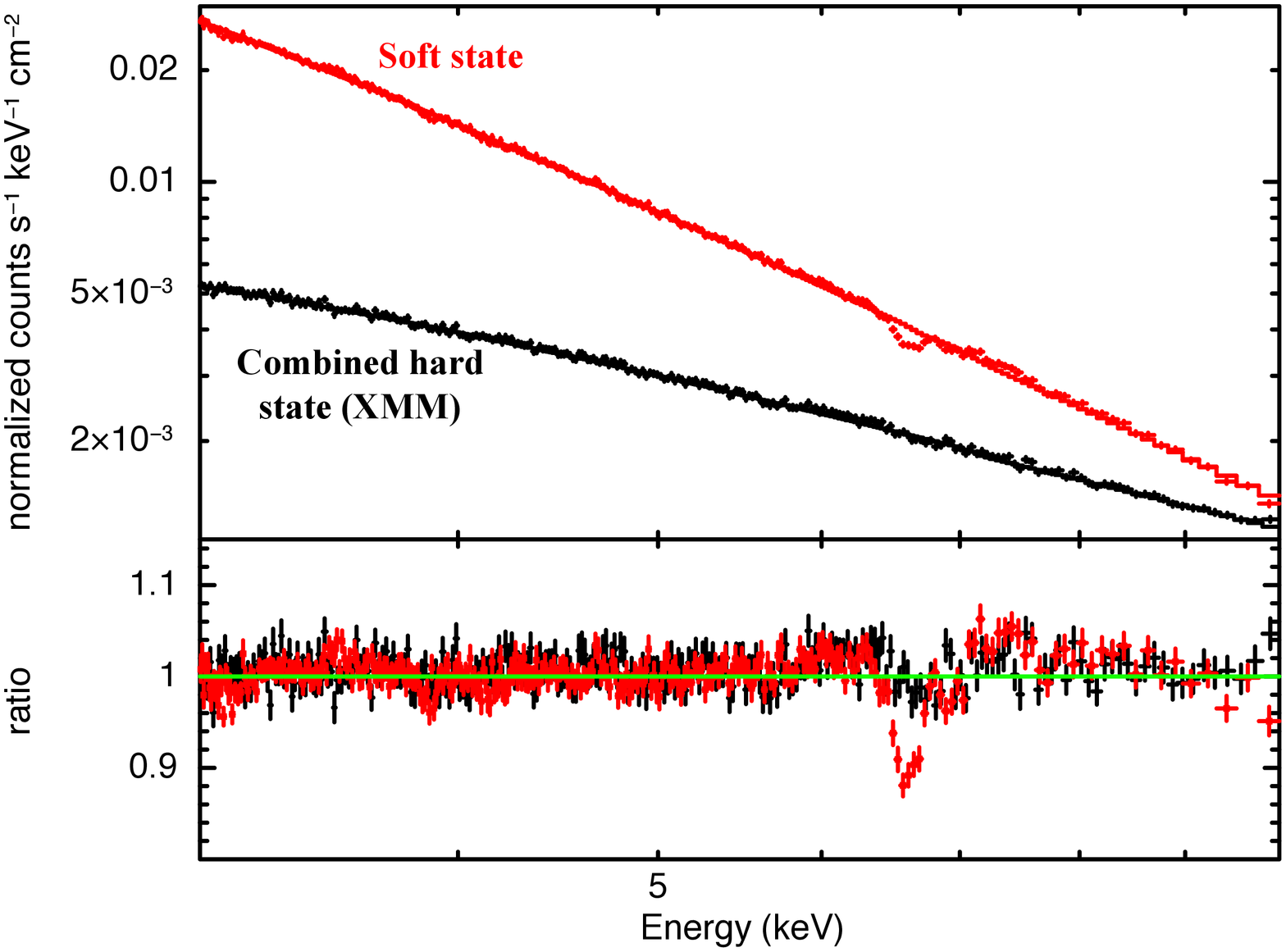}
\caption{Combined spectrum of all the \xmm\ hard state observations (in black)
and soft state observation (in red). The spectra are fitted with a power law 
model absorbed by neutral material. Higher flux and a steeper power law 
slope is observed during the soft state observation, indicating a possible 
contribution for a disc black body component. Very significant Fe~{\sc xxiii-xxv} 
and \Fevs\ absorption lines are present during the soft state observation, 
while they appear absent during the hard states. }
\label{ldsofthard}
\end{figure}

\begin{figure}
\hspace{-1cm}
\includegraphics[width=0.35\textwidth,height=0.58\textwidth,angle=-90]{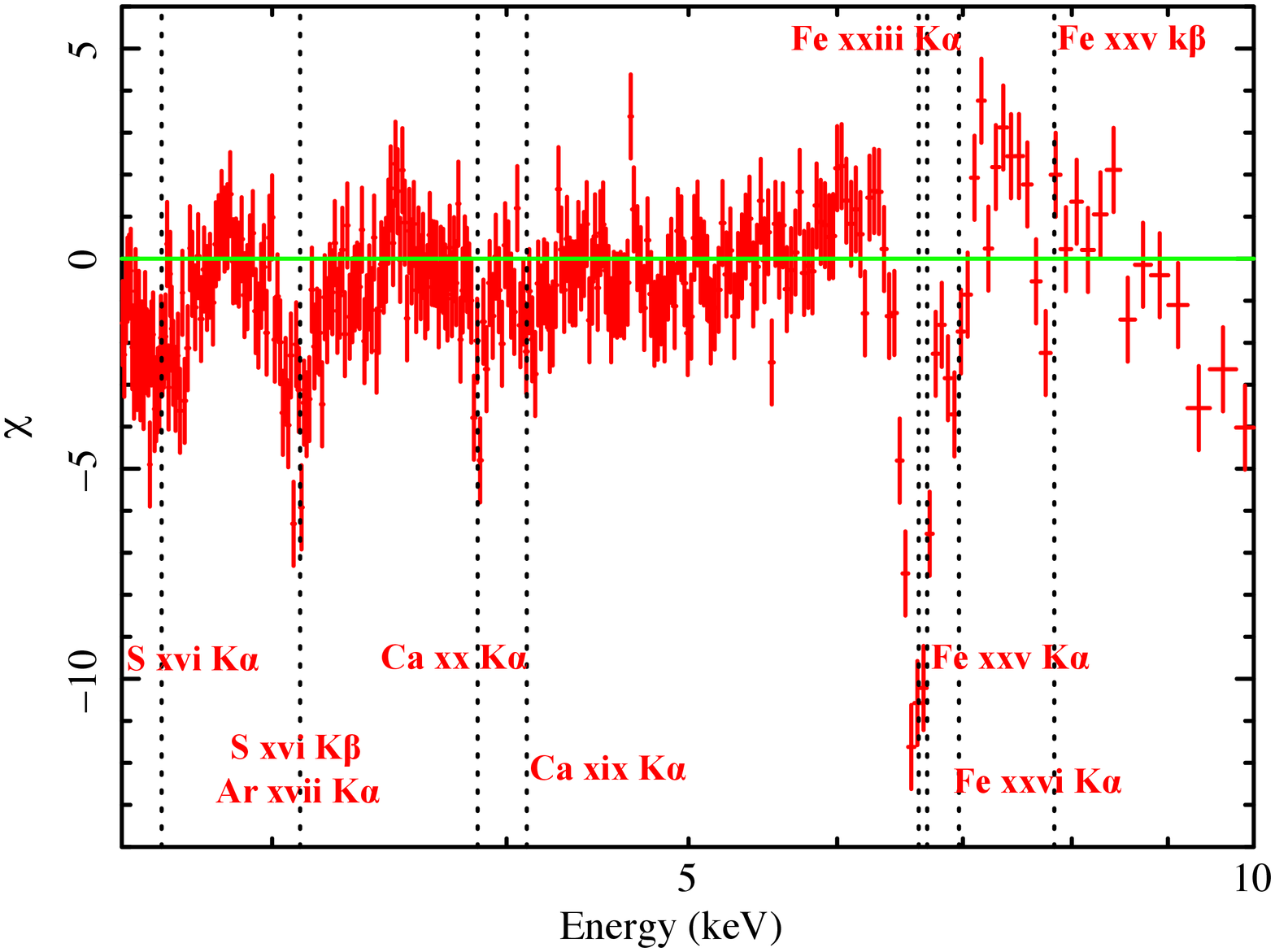}
\caption{Residuals after fitting the 2.5-10~keV band with a power law and 
six absorption lines with Gaussian profile, then removing the absorption lines,
to highlight the significance of the features. The dotted lines indicate the 
energies of the S~{\sc xvi}~K$\alpha$; S~{\sc xvi}~K$\beta$; 
Ca~{\sc xx}~K$\alpha$; Ca~{\sc xix}~K$\alpha$; Fe~{\sc xxiii}~K$\alpha$;
Fe~{\sc xxv}~K$\alpha$; Fe~{\sc xxvi}~K$\alpha$ and 
Fe~{\sc xxv}~K$\beta$ transitions.}
\label{ldrat}
\end{figure}

All the \xmm\ spectra included in this work can be seen in Fig. \ref{ldsofthard}. 
We show in red the source spectrum during the soft state observation (Tab. \ref{data}),
while the combined spectrum of all hard state observations is plotted in black.  
Because we are mainly interested in the Fe~K band and to avoid calibration 
related issues at the 
energies of the edges in the mirrors effective area, we solely fit the data over 
the 2.5-10~keV band. We started by using a power law ({\sc powerlaw} 
in {\sc Xspec}) model absorbed by neutral material ({\sc pha}). This 
resulted in an unacceptable fit ($\chi^2=2277.6$ for 1496 dof), due to  
clear and intense absorption lines that are present at the energies 
of Fe~{\sc xxiii-xxv} and \Fevs\ transitions and at lower energies (see 
Fig. \ref{ldsofthard} and \ref{ldrat}). 
Therefore, we added a Gaussian line that significantly 
improved the fit ($\Delta\chi^2=461.7$ for 3 new free parameters, 
F-test probability $5\times10^{-73}$). 
The line energy is $E=6.63\pm0.01$~keV, with a width 
$\sigma=57\pm18$~eV and an equivalent width $EW=32\pm3$~eV. 
We note that the line energy is formally not consistent with the Fe~{\sc xxv}~K$\alpha$ 
line transition ($E_{\rm \Fevc}=6.7$~keV), suggesting a contribution
by a blend of the rarely observed Fe~{\sc xxiii}~K$\alpha$ and Fe~{\sc xxiv}~K$\alpha$ 
lines\footnote{The energies of the Fe~{\sc xxiii} and Fe~{\sc xxiv} 
transitions with the highest oscillator strength ($f_{ij}$) are: 
$E_{\rm Fe~xxiii}=6.63$ ($f_{ij}=0.67$); 
$E_{\rm Fe~xxiv}=6.65$ (0.15); $E_{\rm Fe~xxiv}=6.66$ (0.47) and
$E_{\rm Fe~xxiv}=6.68$~keV (0.11) (Bianchi et al. 2005).}. 
We then add a second Gaussian line to fit the Fe~{\sc xxvi}~K$\alpha$ 
transition (we assume it has the same width as \Fevc). The fit significantly improves 
($\Delta\chi^2=18.4$ for 2 new parameter, F-test 
probability of $5\times10^{-4}$). The Fe~{\sc xxiii-xxv} and \Fevs\ lines energies 
are now $E=6.63\pm0.01$~keV and $E=6.94\pm0.04$~keV and their 
equivalent widths are $EW_{\rm \Fevc}=32\pm3$~eV and 
$EW_{\rm \Fevs}=8\pm3$~eV, respectively. The lines appear resolved and 
to have a width of $\sigma=45\pm18$~eV. 

Significant residuals are still present around $\sim2.6$, 3.1 and 
$\sim4$~keV (see Fig. \ref{ldrat}). 
Therefore we add other four absorption lines (with width equal to 
the Fe~{\rm xxiii-xxv} one). 
Each of these lines is significant with associated F-test probabilities of 
$8\times10^{-7}$, $2\times10^{-16}$, $4\times10^{-4}$ and $1.9\times10^{-3}$. 
The line energies are: $E=2.603\pm0.015$, $3.119\pm0.015$, 
$3.91\pm0.04$ and $4.18\pm0.04$~keV, thus they are consistent with being 
produced by S~{\sc xvi}~K$\alpha$; a blend of S~{\sc xvi}~K$\beta$ 
and Ar~{\sc xvii}~K$\alpha$; Ca~{\sc xx}~K$\alpha$ and Ca~{\sc xix}~K$\alpha$. 
The line equivalent widths are: $EW=5.9\pm1.7$, $6.1\pm1.2$, 
$3.3\pm1.3$ and $2.9\pm1.3$~eV. 
This phenomenological model composed by a power-law continuum 
absorbed by neutral material plus six Gaussian lines in absorption,  
produces an acceptable fit with $\chi^2=1649.2$ for 
1483 dof. However, we note that an excess of emission is still present 
in the Fe~K band, possibly associated with either a not-modelled ionised 
Fe~K edge or a broad Fe~K emission line.
The best fit power law spectral index is $\Gamma=2.60\pm0.01$. 
The very steep spectral index suggests that an un-modelle d disc black 
body component (as well as, possibly, the boundary layer; e.g. Lin et al. 2007) 
might give a significant contribution at the lowest energies 
considered here. The 3-10 and 6-10~keV observed source fluxes are 
$F_{\rm 3-10keV}=3.9\times10^{-10}$ and 
$F_{\rm 6-10keV}=1.37\times10^{-10}$ erg cm$^{-2}$ s$^{-1}$, respectively. 

The observation of satellite lines due to Fe~{\sc xxiii-xxiv} is quite unusual. 
Alternative hypothesis are that either the line at $E=6.63\pm0.02$~keV is  
associated to the \Fevc\ transition, but produced by inflowing matter, or that 
the energy mismatch might reflect a calibration problem. 
We do observe the same energy of the line ($E=6.63\pm0.02$ keV,
but unresolved $\sigma<70$~eV) in the {\sc mos2} spectrum. 
However, both the EPIC cameras are in timing mode, 
thus they could be affected by the same calibration problems. 
To check if this might be the case, we observe that the ionised 
absorption component fitting the 2.5-10~keV band (with 
$N_{\rm H}=3.0\pm0.5\times10^{22}$ cm$^{-2}$ and $log(\xi)=3.44\pm0.05$; 
see \S 5)
is expected to produce also low energy lines, such as O~{\sc viii} at 
653.6~eV. Indeed, we do detect this line in the RGS spectrum. 
The line energy (after fitting the continuum with a power law model in the 
narrow energy band between 0.6 and 0.7~keV) is E$=653\pm1$~eV and 
unresolved $\sigma<0.8$~eV. No significant red-shift of the line is observed. 
If indeed the O~{\sc viii} line is produced by the same material producing 
the Fe~K lines, this suggests no inflow of the absorbing material with
both a significant contribution from the \Fevc\ inter-combination line plus 
a blend of the Fe~{\sc xxiii} and Fe~{\sc xxiv} lines. 

Finally, to check for any dependence of the ionised absorber with the orbital phase, 
flux level and-or strength of the dipping phenomenon, we divided the light curve 
both in 4 time intervals corresponding to different orbital phases, in 3 flux 
levels and in persistent and dipping intervals and fitted them separately. 
We do not observe any significant variation in the Fe~{\sc xxiii-xxv} and \Fevs\ 
lines (when fitted with Gaussian lines) or in the ionisation parameters 
and/or column density of the photo-ionised absorber (see \S 5). 

\section{Comparison to the hard state} 

The black data in Fig. \ref{ldsofthard} show the combined spectra 
of all the hard state \xmm\ observations. Each hard state spectrum is 
fitted separately by a power law model (normalisation and spectral index 
are left free to vary, see Tab. \ref{data}) absorbed by neutral material ({\sc pha}), 
but with the residuals combined for displaying purposes. 
As previously reported (Diaz-Trigo et al. 2006), no strong Fe~{\sc xxiii-xxv} 
or \Fevs\ absorption line is observed in any of the hard state spectra 
both during the persistent or dipping periods.
In fact, we add to the model two narrow ($\sigma=1$~eV) 
Gaussians to search for the presence of any \Fevc\ or \Fevs\ lines (the 
energies of the Gaussians have been fixed to 6.7 and 6.97~keV,  
the energies of the expected transition). From them, we only find upper limits 
as stringent as $\sim5-15$~eV for both lines in each observation 
(see Tab. \ref{data} and Fig. 1). 

{\it Can this difference be simply due to an ionisation effect?} 
The source persistent (out of dips) luminosity in the 0.5-1~keV band 
is about 5-10 times higher during the soft compared to the hard state 
observations. The luminosity at 7~keV is, instead, less than a factor 
of 2 higher. Therefore, if exactly 
the same material (creating the dips and the high ionisation absorption) 
is along the line of sight during both states, ionisation effect must be at 
play having a significantly higher ionisation parameter during the soft 
state observation. 
Inspecting the light curves, we note that the dipping phenomenon in the 
0.5-5~keV energy band appears less intense during the soft state observation, 
in line with a possible reduction in the opacity induced by increased ionisation
and thus lowering the effect of the dips in the soft X-ray light 
curve\footnote{The dipping phenomenon is known to be erratic, thus this might 
be simply a coincidence, or a consequence of a tilted accretion disc (Diaz-Trigo 
et al. 2009) or the product of the state change. 
In fact, the high inclination line of sight towards \exo\ might be intercepting a 
larger fraction of the disc atmosphere during hard states, characterised by thick 
discs, compared to soft states.}. 
To investigate the effect of the observed luminosity variation on the 
appearance of the Fe~{\sc xxiii-xxv} and \Fevs\ lines, a full characterisation of 
the various absorbing components, taking into account the broad band 
source spectral energy distribution, would be important. 
Such a study is beyond the scope of this paper. However, to have an order 
of magnitude estimate, we have performed a phenomenological fit of 
the soft state spectrum, substituting the six narrow lines with a 
photoionised absorption component (modelled with the {\sc zxipcf} component 
in {\sc Xspec}; Reeves et al. 2008; {\sc zxipcf*pha*powerlaw}). 
This, indeed, provides a reasonable fit ($\chi^2=1745.8$ for 1494 dof), 
suggesting a common origin for these lines. 
We note that this crude 
model is able to reproduce most of the absorption structures (it just leaves
residuals at lower energies compared to the Fe~{\sc xxiii} line, possibly 
due to uncertainties in our knowledge of the oscillator strength for these rarely 
observed transitions) and most of the curvature of the continuum. 
The best fit column density of the ionised layer (assumed 
to be totally covering the X-ray source) is $N_{\rm H}=3.0\pm0.5\times10^{22}$ 
cm$^{-2}$, having an ionisation parameter $log(\xi)=3.44\pm0.05$. 
The column density of the neutral absorber is $N_{\rm H}=8.5\pm0.5\times10^{21}$ 
cm$^{-2}$. Given the narrow energy band used here, this value is not very 
reliable (e.g. Diaz-Trigo et al. 2006 found, fitting the entire energy band, 
$N_{\rm H}=1.1-3.9\times10^{21}$ cm$^{-2}$). 

Using this best fit model as a baseline, we perform a simulation of a hard 
state spectrum. We reduced the ionisation parameter of the {\it zxipcf} 
component by $\sim0.3$ in $log(\xi)$ (as a consequence of the lower 
luminosity at $\sim7$~keV) from $log(\xi)=3.4$ to $log(\xi)=3.1$ and 
simulated the spectrum and then fitted the Fe~K absorption lines. 
We find that even at this lower ionisation parameter and luminosity, 
characteristic of the hard state, two intense Fe~K absorption lines 
at $\sim6.59\pm0.02$~keV and $6.95\pm0.04$, with widths 
$\sigma=0.10\pm0.02$~keV and $\sigma<70$~eV 
and EW of $33\pm4$ and $7\pm3$~eV, should be observed. 
The presence of such lines in the hard state is excluded by our observations. 
This suggests that the presence of Fe~{\sc xxiii-xxv} and \Fevs\ absorption 
lines during the soft state is not {\it simply} due to ionisation effects but 
requires an additional mechanism. However this inference will remain 
tentative until these results will be tested via ad-hoc photo-ionised absorption 
models computed on the basis of the source spectral energy distribution. 

\section{Discussion}

We have observed Fe~K absorption in \exo\ during the soft state, while 
only upper limits are observed during the hard state observations. 
This behaviour resembles what is seen in BH systems (Ponti et al. 2012). 
A previous work by Diaz-Trigo et al. (2006) studied the Fe~K absorption 
features in a sample of dipping-high inclination neutron star systems. 
The authors found that the strongest (with $EW\sim20-40$~eV, such as 
observed here for the first time in \exo) absorption lines in the 
spectra of these systems are \Fevc\ and \Fevs. 
Out of the seven sources in their sample, only \exo\ and 4U~1746-371 
did not show any Fe~K absorption lines. 
The authors suggests that the lack of \Fevc\ and \Fevs\ 
features in \exo\ might be due to a peculiar continuously-dipping 
phenomenon. 
We report here that during the hard state observations (ObsID: 402092010 
and 0160760401) \exo\ clearly shows periods of persistent emission 
with no evidence of dips, but still, no \Fevc\ and \Fevs\ features are observed. 
Alternatively, the authors suggest that for \exo\ the highly ionised 
material is clumpy and located at the outer edge of the accretion disc, 
while being more distributed for the other sources. We investigated 
the dependence of the absorption with orbital phase, but we did 
not find any. 

As \exo\ is a calibration source, more than 20 observations have been 
gathered during the past decade. 
Analysing this wealth of data, we discovered intense Fe~{\sc xxiii-xxv} 
and \Fevs\ absorption lines during the only observation in the soft state.
We note that the soft state observation is characterised by higher soft 
and hard X-ray luminosities and a lowering of the intensity of dipping 
phenomenon in the 0.5-5 keV band. A correlation (anti-correlation) is 
well known between the amplitude of the dips and the absorber 
column density (the absorber ionisation parameter) valid for dipping neutron 
stars (Boirin et al. 2005; Diaz-Trigo et al. 2006). In particular, the persistent 
emission shows the most ionised absorbing component. 
We observe the soft state luminosity ($6-7$~keV) to be higher 
than that of the hard state one (even if by less than a factor of two),  
thus ionisation effects must be at play. However, they do not appear to 
be enough to explain the large variation in ionisation state required to 
generate the Fe~{\sc xxiii-xxv} and \Fevs\ absorption lines from the low 
ionisation material producing the dips. 

We note a remarkable similarity in the properties of high ionisation 
Fe~K absorption in NS and BH systems. They both show Fe~K 
lines with similar equivalent widths ($\sim10-40$~eV), primarily in high 
inclination-dipping sources (Diaz-Trigo et al. 2006; Ponti et al. 2012), 
indicating a similar equatorial geometry. 
However, an important observational difference is related to the 
motion of the absorbers, 
generally observed to be outflowing in BH, while being at rest in several NS. 
Unfortunately, the lack of high resolution grating data during the soft state 
of \exo\ does not allow us to reliably measure the outflow velocity of the Fe~K 
absorption lines, and thus to prove if these features are signatures of a disc 
wind or not. We also note that recent studies of BH systems discovered a 
deep link between the presence of winds and the state of the accretion disc 
(Neilsen \& Lee 2009; Ponti et al. 2012), suggesting that these winds 
are a fundamental ingredient in the accretion process. 
The observation of intense Fe~K absorption features during the soft states 
of \exo\ suggests that the state/wind connection (wind/jet anti-correlation), 
valid for BH, does hold also in this source, and by extension in accreting NS 
in general (Ponti et al. 2014). 

\section*{Acknowledgments}

The authors wish to thank S. Bianchi, M. Guainazzi, M. Freyberg, F. Haberl, T. Dwelly, 
and D. Plant for useful discussion. GP and TMD acknowledge support via an EU Marie 
Curie Intra-European fellowship under contract no. FP-PEOPLE-2012-IEF-331095 and 
FP-PEOPLE-2011-IEF-301355, respectively. This project was funded in part by 
European Research Council Advanced Grant 267697 "4 $\pi$ sky: Extreme Astrophysics 
with Revolutionary Radio Telescopes".

\begin{table*}
\begin{center}
\tiny
\begin{tabular}{ c c c r r c c c c c c c c c c }
\hline
      Obs ID       &   Start date                 & S   &   Exp    & C.E.   & Fil &  Mod &   PP &   Bu / Ecl / Dip / CH / CS / CBack  & EW    & EW  & $\Gamma$            & $F_{\rm 6-10}$ & $F_{\rm 3-6}$ & $F_{\rm 8-10}$ \\
                        &                                   &      &   (ks)    &  (ks)      &      &         &             &                   (c/s)                     & \Fevc\ (eV) &\Fevs\ (eV) &                             &  ($10^{-11}$)     & ($10^{-11}$)   & ($10^{-11}$) \\
                        %&           &                                   &      &             &              &      &         &             &                                               & (eV)   & (eV)  &                             & \\
\hline
\multicolumn{2}{c}{\textbf{\emph{XMM-Newton}} } \\

  0117900901 & 2000-02-23 10:15:09  &  H & 34.0 & 25.1 &  M  &  FW  &   Y  &   8 / 0.6 / 0.27 / 3.6 / 17 / 2.5              & $>-40$      & $>-14$ & $1.53\pm0.12$ & 9.0 & 8.0 & 4.2 \\
  0118700601 & 2000-02-27 07:16:15  &  H & 51.5 & 39.1 &  M  &  FW  &   Y  &   8 / 0.5 / 0.22 / 3.3 / 18 / 1.8          	 & $>-35$      & $>-42$ & $1.45\pm0.13$ & 8.1 & 7.1 & 3.8 \\
  0119710201 & 2000-03-05 20:37:47  &  H & 24.6 & 19.5 &  M  &  LW  &   Y  &   9 / 0.5 / 0.18 / 4.9 / 29 / 2.5            	 & $>-21$      & $>-35$ & $1.43\pm0.09$ & 10.8 & 9.7 & 5.1 \\  
  0123500101 & 2000-04-21 03:10:12  &  H & 18.1 & 11.8 &  M  &  SW  &   N   &   9 / 0.5 / 0.15 / 4.0 / 26 / 0.3             & $>-73$      & $>-43$ & $1.50\pm0.12$ & 7.9 & 7.5 & 3.7 \\ 
  0134561101 & 2001-02-03 17:23:02  &  H &  4.8  &  3.8 &  T  &  FW  &   Y  &  10 / 0.5 / 0.15 / 4.2 / 15 / 1.5         & $>-17\natural$& $>-23$ & $1.26\pm0.05$ & 6.6 & 4.9 &  \\  	  	    	
  0134561201 & 2001-02-04 01:14:50  &  H &  3.0  &  2.2 &  T  &  FW  &   Y  &  10 / 0.5 / 0.2 / 3.9 / 11 / 1.5                 &  "$\natural$& "           & "                        & "     & "     \\
  0134561301 & 2001-02-04 03:47:26  &  H &  3.0  &  2.7 &  T  &  FW  &   Y  &  10 / 0.5 / 0.13 / 3.8 / 19 / 1.5   		 &  "               & "           & "                        & "     & "     \\
  0134561401 & 2001-02-04 07:43:06  &  H &  3.0  &  2.7 &  T  &  FW  &   Y  &  10 / 0.5 / 0.15 / 3.5 / 18 / 1.1   		 &  "               & "           & "                        & "     & "     \\
  0134561501 & 2001-02-04 12:04:21  &  H &  3.0  &  2.6 &  T  &  FW  &   Y  &   8 / 0.5 / 0.16 / 4.0 / 20 / 1.5   		 &  "               & "           & "                        & "     & "     \\
  0160760101 & 2003-09-19 13:30:05  &  H & 88.5 & 56.7 &  M  &  SW  &    N  &   9 / 0.5 / 0.135 / 4.3 / 38 / 0.35                & $>-10$     & $>-10$    & $1.43\pm0.02$ &  8.4 & 7.8 & 4.0 \\
  0160760201 & 2003-09-21 13:31:19  &  H & 90.4 & 59.0 &  M  &  SW  &    N  &   9 / 0.4 / 0.13 / 4.3 / 43 / 0.3   		        & $>-11$      & $>-7.9$   & $1.44\pm0.02$ & 8.5 & 8.0 & 4.0 \\
  0160760301 & 2003-09-23 10:35:20  &  H &107.9 & 66.7 &  M  &  SW  &   N   &  10 / 0.4 / 0.12 / 4.4 / 45 / 0.35   	        & $>-7.3$     & $>-5.5$   & $1.43\pm0.01$ & 8.5 & 8.1 & 4.0 \\
  0160760401 & 2003-09-25 17:22:42  &  H & 73.5 & 48.2 &  M  &  SW  &    N  &   9 / 0.4 / 0.13 / 4.1 / 35 / 0.45   		  & $>-9.0$     & $>-14$   & $1.45\pm0.02$ & 7.8 & 7.4 & 3.7 \\
  0160760601 & 2003-10-21 09:55:39  &  H & 54.9 & 36.2 &  M  &  SW  &    N  &   9 / 0.4 / 0.135 / 4.2 / 40 / 0.4   		  & $>-12.1$   & $>-8.9$  & $1.42\pm0.02$ & 8.7 & 8.1 & 4.1 \\
  0160760801 & 2003-10-25 19:12:54  &  H & 62.4 & 38.7 &  M  &  SW  &    N  &   9.5 / 0.45 / 0.13 / 4.3 / 37 / 0.3   	        & $>-18$      & $>-6.8$   & $1.39\pm0.03$ & 8.6 & 7.9 & 4.1 \\
  0160761301 & 2003-11-12 08:17:18  &  H & 90.7 & 59.3 &  M  &  SW  &    N  &   9 / 0.4 / 0.14 / 4.3 / 35 / 0.3   		        & $>-6.9$     & $>-11$    & $1.41\pm0.02$ & 8.4 & 7.7 & 4.0 \\
  0212480501 & 2005-04-27 13:22:00  &  S & 51.7 & 47.9 & Tck &   T   &    N  &  24 / 1.8 / 0.056\dag / 13.5 / 260 / 0.35    & -32$\pm$3 & 8$\pm$3 & $2.60\pm0.01$ &13.7 & 24.7 & 5.4 \\
  0560180701 & 2008-11-06 08:30:03  &  Q & 27.9 & 23.3 &  M  &  FW  &    N  &  10 / 0.05\ddag / 0.8 / 0.01 / 0.6 / 1.5   \\
  0605560401 & 2009-03-18 00:37:57  &  Q & 41.9 & 34.4 &  T  &  FW  &    N  &  10 / 0.15\ddag / 2.0 / 0.01 / 0.4 / 1.5   \\
  0605560501 & 2009-07-01 05:55:27  &  Q &100.0 & 79.6 &  T  &  FW  &    N  &  10 / 0.1\ddag / 2.0 / 0.01 / 0.38 / 3   \\
  0651690101 & 2010-06-17 05:09:49  &  Q &28.6 & 15.3 &  T  &  FW  &    N  &  10 / 0.1\ddag / 2.0 / 0.01 / 0.4 / 2.0    \\
\multicolumn{2}{c}{\textbf{\emph{Chandra}} } \\
  1017 		   & 2001-04-14 01:13:17  & H &  48.0  &  48.0     &      & HEG &  N       &  -~~$\sharp$                                            & $>-15\flat$  & $>-24$ & & 8.4 & 6.0 & 4.0 \\  
  4573		   & 2003-10-15 08:52:53  & H & 162.9 & 162.9    &      & HEG &  N       &  -                                                          & $>-7.5\flat$ & $>-10$ & & 9.3 & 7.8 & 4.2 \\
  4574		   & 2003-10-13 03:57:32  & H & 123.5 & 123.5   &      & HEG &  N        &  -                                                          & $>-5.8\flat$ & $>-15$ & & 9.0 & 7.6 & 4.1 \\
\multicolumn{2}{c}{\textbf{\emph{Suzaku}} } \\
402092010   & 2007-12-25 05:41:13 & H & 45.9 & 32.7 &    &  1/4   &  N  & 13/0.5/0.15/1.5/11/ -                                            & $>-5.8$         & $>-9.9$ & & 6.6 & 7.3 & 3.9 \\
\hline
\end{tabular}
\caption{List of all the \xmm, \chandra\ and \suzaku\ observations considered. 
In this work we report only the \xmm\ observations with EPIC-pn clean exposures longer than 2~ks. 
The different columns in the table report the observation ID, the observation start date
and time, the source state (H stands for hard; S for soft and Q for quiescent), 
the (EPIC-pn; HETGS; XIS) total and cleaned exposure (after removal, for the EPIC-pn and XIS instruments, 
of periods of increased particle background 
activity, of type I bursts and of periods during eclipses), the EPIC-pn observing filter (T stands for thin; M for medium, 
Tck for thick) and the observing mode (FW stands for Full Frame; SW for Small Window; T for Timing; 1/4 for 
one fourth window), the detection of significant pile up effects. 
The following column shows, in order, the thresholds applied to select bursting, eclipsing and intense dipping 
periods, the hard and soft count rates and the threshold to select out intense particle activity periods. 
The following columns show the \Fevc\ and \Fevs\ lines equivalent widths (or upper limits), the best fit power 
law photon index and the observed fluxes in the $3-6$ and $6-10$~keV bands in units of erg cm$^{-2}$ s$^{-1}$. 
A more exhaustive description of the data reduction and cleaning is provided in \S \ref{xmmDR}. 
\dag During this observation only weak dips are observed. This threshold is used to separate the periods 
of "weak" from "no dips" periods. \ddag During these observations the source is in quiescence. No dip is observed 
and the threshold to select the period during the eclipses has been determined from the soft (0.5-5~keV) band 
light curve. All thresholds and count rates are given in units of counts per seconds. 
$\flat$ Spectrum fitted in the 5-7.7~keV energy range. $\sharp$ We did not apply GTI cuts on the HEG data. 
The detection of narrow absorption features on high resolution HEG data are not expected to be 
significantly affected by the detailed shape of the continuum. Therefore, we apply only the standard cuts 
for increased background activity. $\natural$ We report here the results of the summed spectrum of 
the five observations taken during the  \xmm\ revolution  212. 
}
\label{data}
\end{center}
\end{table*}

\end{document}